\documentstyle[11pt]{article}

\input epsf
\input{tcilatex}
\begin{document}

\hfill CPTH S.650 0998

\hfill September, 1998

\vspace{20pt}

\begin{center}
{\large {\bf \ }}

{\LARGE The N=4 Quantum Conformal Algebra}
\end{center}

\vspace{6pt}

\begin{center}
{\sl Damiano Anselmi}

{\it Centre de Physique Theorique, Ecole Polytechnique, F-91128 Palaiseau
Cedex, FRANCE}~{\sl \footnote{{\sl {\rm {Address starting Oct. 1$^{st}$
1998: Theory Group, CERN, Geneva, Switzerland.}}}}}
\end{center}

\vspace{8pt}

\begin{center}
{\bf Abstract}
\end{center}

We determine the spectrum of currents generated by the operator product
expansion of the energy-momentum tensor in N=4 super-symmetric Yang-Mills
theory. Up to the regular terms and in addition to the multiplet of the
stress tensor, three current
multiplets appear, $\Sigma ,$ $\Xi $ and $\Upsilon ,$ starting with spin 0, 2
and 4, respectively. The OPE's of these new currents
generate an infinite tower of current multiplets, one for each even spin,
which exhibit a universal structure, of length 4 in spin units, 
identified by a two-parameter rational
family. 
Using higher spin techniques developed recently for conformal field
theories, we compute the critical exponents of $\Sigma ,$ $\Xi $ and $%
\Upsilon $ in the $TT$ OPE and prove that the essential 
structure of the algebra holds at arbitrary coupling.
We argue that the algebra closes in the strongly coupled large-$N_c$ limit.
Our results determine the quantum conformal algebra of the
theory and answer several questions that previously remained open.

\vfill\eject

Despite many efforts, low-energy QCD\ is still not well-understood. Recently 
\cite{high}, we planned to approach this problem via a {\it %
detour}, a deeper investigation of 
the conformal
window. Indeed, it seems that this region  
is rich of phenomena that we {\it can} describe 
\cite{noi}, often rigorously, at least in the super-symmetric domain, mainly
because it is not so far as QCD\ from the perturbative region. The idea is
that the lower limit of the conformal window should exhibit at least some 
features of QCD. Something similar does happen in super-symmetric
theories.

The conformal window of asymptotically\ free gauge field theories demands a
better understanding and this perspective turns on the interest for many
related aspects, that are sometimes more abstract, in the sense that
one is lead to consider also theories conceptually far from reality. In the
present paper we study N=4 super-symmetric Yang-Mills theory 
under this perspective, to our knowledge
the simplest family of conformal field theories in four dimensions. 
We apply the ideas of \cite{high}, i.e. techniques similar to those
used in the theory of deep inelastic scattering \cite{muta}. 
This allows us to answer
some problems that remained open in \cite{high} and set up a
basis for future N=1 investigations (particularly in SQCD). 
The hope is that this work
will provide a conceptual basis for the physical problem that we plan to
study and will not make our detour longer than it should be.

Conformal field theories in four dimensions contain many unanswered
questions, although it is true that today we know much more than a couple of
years ago. We feel that the investigation of this subject is only at the
very beginning. It is not necessary to stress once more that there are
enormous differences between conformal field theories in two and four
dimensions. Yet, some similarities persist \cite{noialtri,noi}. We just
mention here the very important role played by higher spin currents in two
dimensional integrable models. Nevertheless, 
in four dimensions it seems that the most
interesting feature of higher spin currents is not their conservation, but
the violation of the conservation condition, in particular the critical
exponent associated with this violation \cite{high}. This exponent is also
the value of the slope of the beta-function associated with the
perturbation generated by the currents \cite{noialtri}.

The plan of the paper is as follows. We start for the analysis of the spin-2
terms of the $TT$ OPE (the spin-0 and spin-1 terms being fully described in 
\cite{noialtri}) and describe the simplifications implied by N=4
supersymmetry on the general OPE\ structure studied in \cite{high}. We
discuss orthogonality of the multiplets and compute their anomalous
dimensions. We then proceed to higher spin and realize that the
structure of the current multiplets is universal, determined by a
simple transformation rule (an SU(4)-invariant bosonic projection of N=4
supersymmetry). This structure encodes the N=4 quantum conformal algebra.
The length of the multiplets is 4 in spin units. The lowest components
(even- and odd-spin) have a common form, while the highest components belong
to a two-parameter rational family. We compute the critical exponents
of the $TT$ OPE, using a technique that
reduces the effort to pure algebra and 
prove that no conserved current other than the stress-tensor
appears. Finally,  we study the phenomena that occur when the 
value of the coupling becomes stronger and
prove that the essential 
structure of the algebra holds at arbitrary coupling.
We argue that the algebra closes in the strongly coupled large-$N_c$ limit.

In \cite{high} it was shown that for spin higher than one there are three
independent even-spin currents (scalar, spinor and vector) and two
independent odd-spin currents (spinor and vector) in the $TT\;$OPE. This
fact determines the length of each current multiplet $\Lambda _{s}$.
It contains three even-spin currents ($\Lambda _{s}$, $\Lambda
_{s+2}$ and $\Lambda _{s+4}$) and two odd-spin currents ($\Lambda _{s+1}$
and $\Lambda _{s+3}$). Only the stress tensor is exceptional, its multiplet
containing a single current. We recall that we focus on $SU(4)$%
-invariant currents, which simplifies the discussion enormously and is
sufficient to our purpose. For a complete discussion of the full N=4
multiplets, in particular the multiplet of the Konishi current, we point out
the recent references \cite{zaffaetal,sezgin}.

We will also discuss certain properties that hold in generality (i.e. in
absence of supersymmetry) and others that hold in the most general
super-symmetric theory. For example, we show that if one of the spin-2
currents appearing at the level of the stress tensor is anomalous
dimensioned, than the other one is also, and that if a higher-spin current
is conserved, than infinitely many are.

The possibility of a higher-spin flavour symmetry in four dimensions is
still open and it is not in contrast with the Coleman-Mandula no-go theorem 
\cite{coleman} that classifies the symmetries of the S-matrix. Indeed, the
higher-spin symmetry would exist 
only at the fixed points and the theorem
does not apply to interacting conformal field theories,
which cannot be described in terms of a theory of scattering.
This is the same reason why one focuses on
the vacuum-to-vacuum amplitudes of conserved currents and their OPE's.

In summary, it could be that higher-spin symmetry is the proper notion to
classify conformal field theories in four dimensions. The N=4 family could
then be identified by the property of having no higher-spin symmetry at all,
the opposite of a free-field theory.

The techniques of \cite{high} are used throughout the paper,
even if often it is not explicitly mentioned. Our investigation is a good
opportunity to see the ideas of \cite{high} at work (in nontrivial conformal
field theories) and also a severe check of the amplitudes of \cite{high}, in
particular the numerical factors of the higher-spin two-point functions.

\vskip 1truecm

{\bf The spin-2 level of the OPE}.

\vskip .3truecm

In super-symmetric theories the number $2N_{F}$ of Majorana fermions equals $%
N_{V}+N_{S}/2$ and therefore it is always possible
to define a spin-2 tensor current $\Xi _{\mu \nu }$ orthogonal to the stress
tensor $T_{\mu \nu }$ in the free-field limit,
with coefficients that do not depend on $N_{V}$ and $%
N_{S}$. The result is unique and reads 
\[
\Xi _{\mu \nu }={\frac{1}{6}}T_{\mu \nu }^{V}-{\frac{2}{3}}T_{\mu \nu
}^{F}+T_{\mu \nu }^{S}. 
\]
Here $T_{\mu \nu }=T_{\mu \nu }^{V}+T_{\mu \nu }^{F}+T_{\mu \nu }^{S}$ is
the stress tensor, equal-weight-sum of its vector, spinor and scalar
contributions. Factorizing out a common factor $\left( \frac{1}{4\pi ^{2}}%
\right) ^{2}{\prod }^{(2)}_{\mu\nu\rho\sigma}\left( {\frac{1}{|x|^{4}}}%
\right) $, one has $\hbox{$<T^V_{\mu\nu}(x)\,T^V_{\rho\sigma}(0)>$}=N_V/5$, $%
<T^F\,T^F>=N_F/10$ and $<T^S\, T^S>=N_S/60$ (indices and spacetime 
points will be
understood from now on).

The orthogonality condition can be preserved off-criticality, in the sense
of perturbation theory, since its implementation is a mere choice of finite
local counter-terms. A second operator, orthogonal to both $T_{\mu \nu }$ and 
$\Xi _{\mu \nu }$ does not exist, in general (with coefficients independent
of $N_{V}$ and $N_{S}$), unless a further symmetry establishes a
relationship between $N_{V}$ and $N_{S}$. This happens for N=4
super-symmetric Yang-Mills theory, where $N_{S}=6N_{V}$ and $%
N_{F}=2N_{V}$. Then the operator 
\[
\Sigma _{\mu \nu }=-T_{\mu \nu }^{V}+\frac{1}{2}T_{\mu \nu }^{F}+T_{\mu \nu
}^{S} 
\]
satisfies the required property.

Orthogonality of $\Xi _{\mu \nu }$ and $T_{\mu \nu }$ means that $\Xi $ can
freely acquire an anomalous dimension $h_{\Xi }$ off-criticality. This
anomalous dimension further evidentiates the ``decoupling'' of $\Xi $ and $T$%
. The same is true for $\Sigma $ in the N=4 theory.

A simple theorem allows us to say that if $h_{\Xi }\neq 0$ then $h_{\Sigma
}\neq 0$ and vice-versa: these two anomalous dimensions are
both zero or both nonzero. The proof uses the OPE projector-invariants
identified in \cite{high}. Let us assume, for example, that $\Sigma _{\mu
\nu }$ is conserved. Then we can work out the $\Sigma _{\mu \nu }(x)~\Sigma
_{\rho \sigma }(y)$ OPE in the same way as for the stress tensor. It is
sufficient to consider the free-field limit. In particular, let us consider
the operator $Q_{\alpha \beta }$ carried by the special invariant SP$_{\mu
\nu ,\rho \sigma ;\alpha \beta }(x-y)$ in the $\Sigma \Sigma $ OPE$.$ Given
that $T^{V,F,S}T^{V,F,S}\rightarrow T^{V,F,S}\times $SP, it is clear that no
linear combination $Q_{\alpha \beta }$ of $T^{V},$ $T^{F}$ and $T^{S}$ other
than the stress tensor gives itself. Moreover, no linear combination $%
Q_{\alpha \beta }$ other than the stress tensor gives a linear combination
of itself and the stress tensor. Consequently a third independent operator
is generated, which has to be conserved, due to the special property of
the projector-invariant SP$_{\mu \nu ,\rho \sigma ;\alpha \beta }(x-y).$
Explicitly, $\Sigma \Sigma \rightarrow $ $(T^{V}+1/4~T^{F}+T^{S})\times $SP
and $T^{V}+1/4~T^{F}+T^{S}=7/10~T-3/14~\Sigma +18/35~\Xi .$

Now, $\Sigma _{\mu \nu }$ belongs to the same current multiplet as the
Konishi current (see below) and therefore its anomalous dimension is
non-vanishing \cite{noialtri}. As a consequence
of the above theorem, $h_{\Xi }\neq 0.$ Moreover,
the explicit computation shows that $h_{\Sigma }\neq h_{\Xi }$, which means
that the $I$-degeneracy of \cite{high} is completely removed. This turns out
to be correct for all terms in the OPE of the N=4 theory.

A more powerful theorem, that we mention here for completeness although it
does not apply to our case, states that if a higher-spin current is
conserved, then infinitely many are. 
The proof is similar to the above one. Suppose that a spin-$2
$ current $\Lambda _{s}$ is conserved, with $s>2,$ and consider the $\Lambda
_{s}\Lambda _{s}$ OPE. There is always a special term SP$_{s}$ in the OPE of
two conserved currents which does not vanish algebraically when tracing it
or taking the divergence, rather it gives a delta function. For $s$ even, SP$%
_{s}$ has dimension four (which produces $\delta (x)$ when tracing and $%
\partial \delta (x)$ when taking the divergence), while for $s$ odd SP$_{s}$
has dimension three (like in the case of the OPE\ of two conserved spin-1
currents). Now, since $\Lambda _{s}$ has dimension $2+s$, the operator
carried by SP$_{s}$ has dimension $2s$ (and therefore spin $2s-2)$ when $s$
is even, while it has spin $2s-1$ when $s$ is odd. Therefore an $s$ greater
than two generates a infinite tower of conserved currents.

This property just proved is simply the known fact, here
derived at the quantum level, that any higher spin symmetry algebra is
infinite dimensional. It is somewhat opposite to the complete removal of
the $I$-degeneracy (but there could be higher-spin symmetry with no more
than one conserved current for each spin).

The form of the N=4 OPE reads (see \cite{high} for the notation)
\begin{equation}
T_{\mu \nu }(x)~T_{\rho \sigma }(y) =\frac{1}{60}~\left( \frac{1}{4\pi ^{2}%
}\right) ^{2}~{\prod }_{\mu \nu ,\rho \sigma }^{(2)}\left( \frac{c_{2}}{%
|x-y|^{4}}\right) +{\frac{1}{4\pi ^{2}}}~T_{\alpha \beta }{\rm ~}\widetilde{%
{\rm SP}}_{\mu \nu ,\rho \sigma ;\alpha \beta }(x-y),
\end{equation}
plus anomalous dimensioned operators, descendants
and regular terms. Here $c_2=30 N_V$.
The invariant $\widetilde{{\rm SP}}_{\mu \nu ,\rho \sigma ;\alpha
\beta }(x-y)$ multiplying the stress-tensor reads
\begin{eqnarray*}
\widetilde{{\rm SP}}_{\mu \nu ,\rho \sigma ;\alpha \beta }(x-y) &=&{\rm SP}%
_{\mu \nu ,\rho \sigma ;\alpha \beta }(x-y)+\frac{43}{480}~{\prod }_{\mu \nu
,\rho \sigma }^{(2)}\partial _{\alpha }\partial _{\beta }\left( |x-y|^{2}\ln
|x-y|^{2}M^{2}\right)  \\
&&-\frac{5}{32}~{\prod }_{\mu \nu \alpha ,\beta \rho \sigma }^{(3)}\left(
|x-y|^{2}\ln |x-y|^{2}M^{2}\right) .
\end{eqnarray*}
The importance of this invariant is related to
its role in the $<T(x)T(y)T(z)>$  three-point function. Since 
$\Sigma_{\mu\nu}$ and $\Xi_{\mu\nu}$
are orthogonal to $T$, only this invariant contributes
to the three-point functions in this limit. The full correlator is
reconstructed uniquely by the invariant $\widetilde{{\rm SP}}$ and its
descendants. Now, the three-point function, and therefore also $\widetilde{%
{\rm SP}}$, encodes both the quantity $a$ and $c$ \cite{noi}
and the nonrenormalization
theorem says that $\widetilde{{\rm SP}}$ is independent of the coupling
constant.

We now begin the analysis of the dimensioned operators. The
Lagrangian of the theory is \cite{Sohnius}\footnotemark 
\footnotetext{
In our formulas we convert the Minkowskian notation of Sohnius \cite{Sohnius}
to the Euclidean framework by the substitutions $\delta_{\mu\nu}\rightarrow-%
\delta_{\mu\nu}$, $T^{V,F,S}\rightarrow -T^{V,F,S}$, $\varepsilon_{\mu\nu\rho%
\sigma} \rightarrow -i\varepsilon_{\mu\nu\rho\sigma}$ and $\gamma_\mu,
\gamma_5\rightarrow -i\gamma_\mu,-i\gamma_5$. Moreover, we multiply $A$ 
and $B$ by a factor 2.} 
\[
{\cal L}={\rm tr}\left[ \frac{1}{4}F^{2}+\frac{1}{2}\bar{\lambda}_{i}\gamma
_{\mu }\nabla _{\mu }\lambda _{i}+\frac{1}{2}\left( \nabla _{\mu
}A_{ij}\right) ^{2}+\frac{1}{2}\left( \nabla _{\mu }B_{ij}\right)
^{2}+...\right].
\]
The scalar operator appearing in the $TT$ OPE is 
\[
\Sigma_0 =A_{ij}^{2}+B_{ij}^{2}.
\]
The
axial current appearing in the $TT$ OPE is the $SU(4)$ singlet current that
belongs to the same super-field as the scalar operator $\Sigma_0 $, 
\[
\Sigma _{\mu }={\frac{1}{2}}\bar{\lambda}_{i}\gamma _{5}\gamma _{\mu
}\lambda _{i}.
\]
The factor $1/2$ w.r.t. \cite{high} is due to the Majorana condition. In
particular, applying two super-symmetries $\delta _{\zeta }$ to $\Sigma_0 $
with the same fermionic parameter $\zeta $ and looking at the coefficient of 
$\bar{\zeta}_{i}\gamma _{\mu }\gamma _{5}\zeta _{j},$ one obtains the full
set of Konishi currents $\Sigma _{\mu }^{ij}$ in the ${\bf 4}\otimes {\bf 
\bar{4}=15\oplus 1}$ representations of $SU(4)$, 
\[
\Sigma _{\mu }^{ij}=\frac{1}{2}(\bar{\lambda}_{i}\gamma _{5}\gamma _{\mu
}\lambda _{j}-\delta _{ij}\bar{\lambda}_{k}\gamma _{5}\gamma _{\mu }\lambda
_{k})+4iA_{ik}\overleftrightarrow{\nabla }_{\mu }B_{kj}.
\]
The trace $\Sigma _{\mu }$ is the spin-1 operator appearing in the $TT$ OPE.
Instead, isolating the coefficient of $\bar{\zeta}_{i}\gamma _{\mu }\zeta
_{j}$ in $\delta _{\zeta }^{2}\Sigma_0 ,$ one gets the set of vector currents 
\[
V_{\mu }^{ij}=-i\bar{\lambda}_{i}\gamma _{\mu }\lambda _{j}+4iA_{ik}%
\overleftrightarrow{\nabla }_{\mu }A_{kj}+4iB_{ik}\overleftrightarrow{\nabla }%
_{\mu }B_{kj}.
\]
Now, applying the procedure again to $\Sigma _{\mu }$ one gets precisely the
operator $\Sigma _{\mu \nu }$, which is the only $SU(4)$ spin-2 current
related to $\Sigma_0 $ by supersymmetry. Applying the same procedure to $%
\Sigma _{\mu }^{ij}$ and $V_{\mu }^{ij}$ non new $SU(4)$ singlet is found.
We conclude that $\Sigma_0 $, $\Sigma _{\mu }$ and $\Sigma _{\mu \nu }$ have
the same anomalous dimension, which is for $G=SU(N_{c})$ \cite{noialtri} 
\[
h_{\Sigma }=3N_{c}{\frac{\alpha }{\pi }}
\]
($\alpha =g^{2}/(4\pi )$ and $N_{c}\rightarrow C(G)$ for a generic gauge
group). $\Xi _{\mu \nu }$ is not related to $\Sigma $ and so in principle it
can have a different anomalous dimension. Nevertheless we know a priori that
it is non-zero.

We compute $h_\Xi$ by combining several arguments of \cite{high}. In
particular, we know that conformality and tracelessness implies 
\[
<{\cal J}^{(i)}_{\mu\nu}(x)\,{\cal J}^{(j)}_{\rho\sigma}(0)>= {\frac{8}{%
3(4\pi^2)^2}}{\frac{c^{(2)}_i(g)\delta_{ij} }{(|x|\mu)^{2h_i(g)}}} {\frac{{\cal I}%
_{\mu\nu,\rho\sigma}(x)}{|x|^8}}, 
\]
where $\{{\cal J}_2^{(i)}\}=\{T,\Sigma_2,\Xi_2\}$, 
$i=T,\Sigma,\Xi$ and $c^{(2)}_i(g)$ are the spin-2 central charges,
their free-field values being
$\{c^{(2)}_i(0)\}=\{30\,N_V,21\,N_V,35/3\, N_V\}$. Moreover,
${\cal I}%
_{\mu\nu,\rho\sigma}={\cal I}_{\mu\rho}{\cal I}_{\nu\sigma}+ {\cal I}%
_{\mu\sigma}{\cal I}_{\nu\rho}-{\frac{1}{2}}\delta_{\mu\nu}\delta_{\rho%
\sigma}$ and ${\cal I}_{\mu\nu}=\delta_{\mu\nu}-2x_\mu x_\nu/|x|^2$ as
usual. This implies in turn that 
\begin{equation}
<\partial_\mu{\cal J}^{(i)}_{\mu\nu}(x)\, \partial_\rho{\cal J}%
^{(j)}_{\rho\sigma}(0)>= h_i{\frac{3}{2\pi^4}}c^{(2)}_i(0)\,\delta_{ij} {\frac{%
{\cal I}_{\nu\sigma}(x)}{|x|^{10}}}+{\cal O}(g^4). 
\label{ty}
\end{equation}
This formula exhibits the violation of the
conservation condition. Now, given that $\partial_\mu{\cal J}%
^{(i)}_{\mu\nu}= {\cal O}(g)$ and $h_i={\cal O}(g^2)$, we see that we can
neglect ${\cal O}(g^2)$ in $\partial_\mu{\cal J}^{(i)}_{\mu\nu}$ and retain
only the Yukawa and gauge terms. Moreover, since the matrix of renormalization
constants $Z_{ij}$ has the form $\delta_{ij}+{\cal O}(g^2)$, we can neglect
the difference between bare and renormalized operators at this level. The
same can be said about any finite factor attached to ${\cal J}^{(i)}$ by its
very definition (we recall that by definition the expression of the higher spin
non-conserved currents is read inside the $TT$ OPE \cite
{high}, which fixes their finite factors also), as well as about anomalous
terms (which we expect to be present, since they are present in the
divergence of the Konishi current $\Sigma_\mu$; nevertheless they begin at
least with the order $g^2$). In practice this method reduces the computation
of the lowest order of $h_i$ to pure algebra.

The procedure is completely general,
if one excludes the scalar operator $\Sigma_0$ for which a two-loop graph
computation is necessary (done in \cite{noialtri}). We can check the method
with $\Sigma_\mu$. We have 
\[
<\Sigma_\mu(x)\,\Sigma_\nu(0)>=-{\frac{2N_V}{\pi^4}} {\frac{{\cal I}%
_{\mu\nu}(x)}{|x|^{6+2h_\Sigma}}}+{\rm irrelevant}, \quad
<\partial\cdot\Sigma(x)\,\,\partial\cdot\Sigma(0)>= -h_\Sigma {\frac{16N_V}{%
\pi^4}} {\frac{1}{|x|^8}}+{\rm irr}. 
\]
The field equations give $\partial\cdot
\Sigma=2igf_{abc}\bar\lambda_i^a\gamma_5 (A_{ij}^b-i\gamma_5 B^b_{ij})
\lambda_j^c$ and $<\partial\Sigma\,\partial\Sigma>$ gives immediately $%
h_\Sigma=3N_c\, \alpha/\pi$.

The spin-2 case is more subtle. We first covariantize the free-field
operators so as to preserve formal conformality, i.e. conformality at the
classical level. Then we extract all traces. A nontrivial check that this
procedure is good is the emergence of the correct stress-tensor, even if we
do not apply directly the Noether method. Our method is more general and
works also for non-conserved currents. At higher orders the situation is
further complicated by the renormalization constants $Z_{ij}$ and the mixing
with descendants of the improvement terms \cite{high}.

Conformality at the classical level is implemented by the transformations 
\begin{eqnarray}
A_{ij} &\rightarrow &|x|^{2}A_{ij}\qquad B_{ij}\rightarrow
|x|^{2}B_{ij},\qquad \lambda _{i}\rightarrow |x|^{2}x\!\!\!\slash\gamma
_{5}\lambda _{i},\qquad \bar{\lambda}_{i}\rightarrow |x|^{2}\bar{\lambda}%
_{i}\gamma _{5}x\!\!\!\slash,  \nonumber \\
\partial _{\mu } &\rightarrow &|x|^{2}{\cal I}_{\mu \nu }\partial _{\nu
},\qquad D_{\mu }\rightarrow |x|^{2}{\cal I}_{\mu \nu }D_{\nu },\qquad
F_{\mu \nu }\rightarrow |x|^{4}{\cal I}_{\mu \rho }{\cal I}_{\nu \sigma
}F_{\rho \sigma },  \label{pus}
\end{eqnarray}
and selects the ``improved'' currents \cite{high}. The transformation of $A_{\mu }$ 
(as well as $D_{\mu }$ and $F_{\mu \nu }$ in the non-Abelian case) 
is actually
defined only up to gauge-transformations, therefore it is understood that
the above rules have to be applied to gauge-invariant operators. 
Formal conformality goes through the
covariantization process with no problem and follows exactly
as for $g=0$ (with $\partial\rightarrow D$).
The double covariant
derivatives do not present any ambiguity,
since their indices are always symmetrized.
Moreover, they can often be moved away by 
using the improvement terms (see below).
Classical conformality is
sufficient for our purposes since the anomalous dimension $h$ is order $g^{2}
$ and its effects can be neglected here.

Therefore we define 
\begin{eqnarray}
\Xi _{\mu \nu }&=&{\frac{1}{6}} T_{\mu \nu }^{V}-{\frac{2 }{3}}T_{\mu \nu
}^{F}+T_{\mu \nu }^{S}-{\frac{1}{4}} \delta_{\mu\nu}\left({\frac{1}{6}}%
\Theta^V-{\frac{2}{3}}\Theta^F+\Theta^S\right),  \nonumber \\
\Sigma _{\mu \nu }&=&-T_{\mu \nu }^{V}+\frac{1}{2}T_{\mu \nu }^{F}+T_{\mu
\nu }^{S}-{\frac{1}{4}} \delta_{\mu\nu}\left(-\Theta^V+{\frac{1}{2}}%
\Theta^F+\Theta^S\right),  \nonumber \\
T_{\mu\nu}&=&T_{\mu \nu }^{V}+T_{\mu \nu }^{F}+T_{\mu \nu}^{S}-{\frac{1}{4}}
\delta_{\mu\nu}\left(\Theta^V+\Theta^F+\Theta^S\right),  \nonumber
\end{eqnarray}
where $\Theta$ denotes the traces and 
\begin{eqnarray}
T^S_{\mu\nu}&=&D_\mu A^a_{ij}D_\nu A^a_{ij}-{\frac{1}{2}}
\delta_{\mu\nu}(D_\alpha A^a_{ij})^2+ D_\mu B^a_{ij}D_\nu B^a_{ij}-{\frac{1}{%
2}} \delta_{\mu\nu}(D_\alpha B^a_{ij})^2 -{\frac{1}{6}}\pi_{\mu\nu}(A^2+B^2)
,  \nonumber \\
T^F_{\mu\nu}&=&{\frac{1}{4}}(\bar\lambda_i^a \gamma_\mu D_\nu \lambda_i^a+
\bar\lambda_i^a \gamma_\nu D_\mu \lambda_i^a),\qquad\qquad\qquad
T^V_{\mu\nu}=F^a_{\mu\rho}F^a_{\nu\rho}-{\frac{1}{4}}\delta_{\mu\nu} F^2,
\nonumber
\end{eqnarray}
where $\pi_{\mu\nu}=\partial_\mu\partial_\nu-\Box\delta_{\mu\nu}$.
The calculation is a bit lengthy, but straightforward. In this illustrative
case, we give the details
and in the rest of the paper we present the
derivation of our results more schematically. The divergences $%
\partial_\mu{\cal J}^{(i)}_{\mu\nu}$ are the sums of three operators $u_\mu$%
, $v_\mu$ and $z_\mu$ that are orthogonal to one another
at the free-field level.
Precisely, 
\begin{eqnarray}
u_\mu&=&{\frac{g}{2}}f_{abc}\bar\lambda_i^a\gamma_\nu\lambda_i^b
F^c_{\mu\nu}, \qquad\qquad\quad v_\mu=-gf_{abc}(A_{ij}^a%
\partial_\nu A_{ij}^b+B_{ij}^a\partial_\nu B^b_{ij})F^c_{\mu\nu},  \nonumber
\\
z_\mu&=&-gf_{abc}\bar\lambda_i^a(\partial_\mu A^b_{ij}-i
\gamma_5\partial_\mu B_{ij}^b)\lambda_j^c +{\frac{g}{4}}f_{abc}\partial_\mu[%
\bar\lambda_i^a(A_{ij}^b-i\gamma_5 B_{ij}^b) \lambda_j^c].  \nonumber
\end{eqnarray}
Actually, only two combinations of these three operators appear in $%
\partial_\mu{\cal J}^{(i)}_{\mu\nu}$, since 
$T_{\mu\nu}$ is conserved.
The other two divergences are 
\begin{equation}
\partial_\nu\Sigma_{\mu\nu}={\frac{3}{2}}u_\mu-v_\mu-{\frac{1}{2}}%
z_\mu,\qquad\qquad \partial_\nu\Xi_{\mu\nu}=-{\frac{5}{3}}\left({\frac{1}{2}}%
u_\mu+ {\frac{1}{4}}v_\mu+z_\mu\right).
\end{equation}
Factorizing out the common factor $g^2 N_cN_V {\cal I}_{\mu\nu}(x)/(%
\pi^6|x|^{10})$, we have $<uu>=6$, $<vv>=9$ and $<zz>=9/2$, which gives 
\begin{equation}
h_\Sigma=3N_c{\frac{\alpha}{\pi}},\qquad\qquad\quad h_\Xi={\frac{25}{6}}N_c {%
\frac{\alpha}{\pi}},  \label{h}
\end{equation}
and, as a further nontrivial check, $<\partial_\mu\Xi_{\mu\nu}(x)\,\partial_%
\rho\Sigma_{\rho\sigma}(0)>=0$.

As expected, $h_\Xi$ is nonzero and different from $h_\Sigma$, stressing
that $\Xi$ belongs to a new super-field and that the $I$-degeneracy
is completely removed.

\vskip 1truecm

{\bf The spin-3 level of the OPE}.

\vskip .3truecm

Now, we turn to the spin-3 currents. As usual, we apply two supersymmetry
transformations $\delta _{\zeta }$ to the spin-2 currents, look at the
coefficient of $\bar{\zeta}_{i}\gamma _{\mu }\gamma _{5}\zeta _{j}$ and then
trace it in $i,j$. One obtains two linear combinations of the spin-3
currents appearing in the $TT$ OPE \cite{high} (vector and
spinor). $\delta _{\zeta }^{2}T_{\mu \nu }$ gives zero, while the
other combinations are 
\begin{equation}
\Sigma _{\mu \nu }\rightarrow \Sigma _{\mu \nu \rho }={\cal A}_{\mu \nu \rho
}^{F}-8{\cal A}_{\mu \nu \rho }^{V},\qquad \Xi _{\mu \nu }\rightarrow \Xi
_{\mu \nu \rho }={\cal A}_{\mu \nu \rho }^{F}+{\frac{16}{5}}{\cal A}_{\mu
\nu \rho }^{V},
\end{equation}
where 
\begin{eqnarray}
{\cal A}_{\mu \nu \rho }^{F} &=&\sum_{{\rm symm}}-2D_{\mu }\bar{\lambda}%
_{i}^{a}\gamma _{5}\gamma _{\rho }D_{\nu }\lambda _{i}^{a}+\frac{2}{5}
\partial _{\mu }\partial _{\nu } \left( \bar{\lambda}_{i}^{a}\gamma
_{5}\gamma _{\rho }\lambda _{i}^{a}\right) -{\rm traces},  \nonumber \\
{\cal A}_{\mu \nu \rho }^{V} &=&\frac{1}{3}\left[ F_{\nu \alpha }^{+}%
\overleftrightarrow{D_{\mu }}F_{\alpha \rho }^{-}+F_{\mu \alpha }^{+}%
\overleftrightarrow{D_{\nu }}F_{\alpha \rho }^{-}+F_{\mu \alpha }^{+}%
\overleftrightarrow{D_{\rho }}F_{\alpha \nu }^{-}\right] -{\rm traces}.
\end{eqnarray}
We have already written the expressions for non-vanishing $g$. 
It is somewhat convenient to move the double covariant derivatives
to the improvement terms, where they become simple derivatives. We
shrink the notation by denoting the spin of an element in the current
multiplet with a subscript and suppressing indices when possible. So, we
write $\Sigma _{0},$ $\Sigma _{3},$ $\Xi _{2},$ etcetera.

The first thing to check is orthogonality between $\Sigma _{\mu \nu \rho }$
and $\Xi _{\alpha \beta \gamma }$. Factorizing out the common factor $%
N_{V}\left( \frac{1}{4\pi ^{2}}\right) ^{2}{\prod }_{\mu \nu \rho ,\alpha
\beta \gamma }^{(3)}\left( {\frac{1}{|x|^{4}}}\right) $ and keeping into
account of the Majorana condition for the spinors $\lambda _{i}$, the
results of \cite{high} give $<{\cal A}_{3}^{F}~{\cal A}_{3}^{F}>=32/105$ and 
$<{\cal A}_{3}^{V}~{\cal A}_{3}^{V}>=1/84$, where-from $<\Sigma _{3}~\Xi
_{3}>=0$ follows. This is a good check of the numerical factors of \cite
{high}. One then finds $<\Sigma _{3}~\Sigma _{3}>=16/15$ and $<\Xi _{3}~\Xi
_{3}>=32/75$.

Now, lets us write the generic form of the spin-3 two-point function, 
\[
<{\cal A}_{\mu \nu \rho }^{(i)}(x)\,{\cal A}_{\alpha \beta \gamma
}^{(j)}(0)>=\left( \frac{1}{4\pi ^{2}}\right) ^{2}{\frac{c_{i}^{(3)}\delta
_{ij}}{(|x|\mu )^{2h_{i}}}}{\prod }_{\mu \nu \rho ,\alpha \beta \gamma
}^{(3)}\left( {\frac{1}{|x|^{4}}}\right) =c_{i}^{(3)}\delta _{ij}\left( 
\frac{1}{4\pi ^{2}}\right) ^{2}{\frac{{\cal I}_{\mu \nu \rho ,\alpha \beta
\gamma }^{(3)}(x)}{(|x|\mu )^{2h_{i}}|x|^{10}}}, 
\]
where $\{{\cal A}_3^{(i)}\}=\{\Sigma_3 ,\Xi_3 \}$, $i=\Sigma,\Xi$ and $%
\{c_{i}^{(3)}\}=\{16N_{V}/15,32N_{V}/75\}$. The correlator of the two
divergences is 
\begin{equation}
<\partial _{\rho }{\cal A}_{\mu \nu \rho }^{(i)}(x)\,\partial _{\gamma }%
{\cal A}_{\alpha \beta \gamma }^{(j)}(0)>=-{\frac{896}{5}}h_{i}\left( \frac{1%
}{4\pi ^{2}}\right) ^{2}{\frac{c_{i}^{(3)}\delta _{ij}}{|x|^{4}}}{\prod }%
_{\mu \nu ,\alpha \beta }^{(2)}\left( {\frac{1}{|x|^{4}}}\right) .
\label{div3}
\end{equation}

Using this formula we 
check the value of $h_{\Xi }$. We can restrict ourselves to a single current
and, to simplify further the calculation (and to prepare for a new, more
difficult one to be done below) we calculate the two-point function of the
divergence of the spin-3 vector current 
\[
{\cal A}_{3}^{V}={\frac{5}{56}}(\Xi _{3}-\Sigma _{3}). 
\]
Then (\ref{div3}) gives 
\[
<\partial _{\rho }{\cal A}_{\mu \nu \rho }^{V}(x)\,\partial _{\gamma }{\cal A%
}_{\alpha \beta \gamma }^{V}(0)>=-{\frac{10}{7}}\left( {\frac{1}{4\pi ^{2}}}%
\right) ^{2}{\frac{(c^{(3)}_{\Sigma }h_{\Sigma }+c^{(3)}_{\Xi }h_{\Xi })}{|x|^{4}}}{%
\prod }_{\mu \nu ,\alpha \beta }^{(2)}\left( {\frac{1}{|x|^{4}}}\right) , 
\]
The explicit computation gives indeed 
\[
<\partial _{\rho }{\cal A}_{\mu \nu \rho }^{V}(x)\,\partial _{\gamma }{\cal A%
}_{\alpha \beta \gamma }^{V}(0)>=-{\frac{64}{9}}N_{c}N_{V}\left( \frac{1}{%
4\pi ^{2}}\right) ^{2}{\frac{\alpha }{\pi }}{\frac{1}{|x|^{4}}}{\prod }_{\mu
\nu ,\alpha \beta }^{(2)}\left( {\frac{1}{|x|^{4}}}\right) , 
\]
which perfectly agrees with the values (\ref{h})
of $h_\Sigma$ and $h_\Xi$. When performing
this computation the following observation
helps reducing the effort and improves the precision.
The field equations show that the contracted derivatives $E_{\mu
}=D_{\nu }F_{\nu \mu }^{\pm }$ are orthogonal to $F^{\pm }$ (since $E_{\mu }$
contain only gauginos $\lambda _{i}$ and scalar fields $A,B$).
Their two-point function is $%
<E_{\mu }^{a}(x)~E_{\nu }^{b}(0)>=11g^{2}N_{c}\delta ^{ab}/(4\pi ^{2})^{2}~%
{\cal I}_{\mu \nu }(x)/|x|^{6}$. Therefore, the divergence 
$\partial _{\rho }{\cal A}_{\mu \nu \rho }^{V}$ is the sum of two terms
that are separately conformal: one term contains $E_{\mu }^{a}$;
the other term is cubic in the field strength and comes
from the commutation of the covariant derivatives. The field equations 
can be freely used, since, as we know, anomalous contributions
are negligible at this level.

\vskip 1truecm

{\bf The spin-4 and higher levels of the OPE}.

\vskip .3truecm

In order to proceed, we observe that the operation $\delta _{\zeta }^{2}$
that relates the currents of a multiplet is actually universal. To
see this, let us normalize the currents as follows, 
\begin{eqnarray*}
{\cal J}^{V} &=&F_{\mu \alpha }^{+}\overleftrightarrow{\Omega }_{{\rm even}%
}F_{\alpha \nu }^{-}+{\rm impr}.,\quad \quad {\cal J}^{F}=\frac{1}{2}\bar{%
\lambda}_{i}\gamma _{\mu }\overleftrightarrow{\Omega }_{{\rm odd}}\lambda
_{i}+{\rm impr}., \\
{\cal A}^{V} &=&F_{\mu \alpha }^{+}\overleftrightarrow{\Omega }_{{\rm odd}%
}F_{\alpha \nu }^{-}+{\rm impr}.,\quad \quad {\cal A}^{F}=\frac{1}{2}\bar{%
\lambda}_{i}\gamma _{5}\gamma _{\mu }\overleftrightarrow{\Omega }_{{\rm even}%
}\lambda _{i}+{\rm impr}., \\
{\cal J}^{S} &=&A_{ij}\overleftrightarrow{\Omega }_{{\rm even}%
}A_{ij}+B_{ij}\overleftrightarrow{\Omega }_{{\rm even}}B_{ij}+%
{\rm impr}.,
\end{eqnarray*}
where $\overleftrightarrow{\Omega }_{{\rm even/odd}}$ denotes an even/odd
string of derivative operators $\overleftrightarrow{\partial }$ and
``impr.'' stands for the improvement terms \cite{high}
(here they are not written
explicitly, but they are important for all our computations). This condensed
notation is particularly useful, since the operation $\delta _{\zeta }^{2}$
commutes with $\overleftrightarrow{\Omega }$. Therefore, a simple 
set of basic
rules suffices to determine the operation $\delta _{\zeta
}^{2} $. The result is 
\begin{eqnarray*}
{\cal A}_{2s-1}^{V} &\rightarrow &-4~{\cal J}_{2s}^{V}+\frac{1}{4}~{\cal J}%
_{2s}^{F},\qquad \qquad {\cal A}_{2s-1}^{F}\rightarrow -16~{\cal J}%
_{2s}^{V}-2~{\cal J}_{2s}^{F}+2~{\cal J}_{2s}^{S}, \\
{\cal J}_{2s}^{V} &\rightarrow &-4~{\cal A}_{2s+1}^{V}+\frac{1}{4}~{\cal A}%
_{2s+1}^{F},\qquad \qquad {\cal J}_{2s}^{F}\rightarrow -16~{\cal A}%
_{2s+1}^{V}-2~{\cal A}_{2s+1}^{F}, \\
{\cal J}_{2s}^{S} &\rightarrow &-6~{\cal A}_{2s+1}^{F},
\end{eqnarray*}
independently of the spin. The arrow raises the spin
by one unit. In particular, we rescale $\Sigma_2$ and $\Xi_2$ by a factor -4.
We normalize
currents in the current multiplets by fixing the coefficients 
of ${\cal J}^{S}$ and
${\cal A}^{F}$ to one, for even and odd
spin respectively.
Note that for spin two we have $T^{V}=-2{\cal J}_2^{V},$ $T^{F}=%
{\cal J}_2^{F}/2$ and $T^{S}=-{\cal J}_2^{S}/4$.

Let us analyse the four current multiplets of the $TT$ OPE. The shortest
multiplet is the one of the stress tensor, which contains only the
stress tensor. The other three current multiplets 
before the regular terms (the Konishi multiplet $\Sigma $, the $\Xi $%
-multiplet and a new multiplet $\Upsilon$) are
\[
\begin{tabular}{lllll}
$\Sigma _{0}={\cal J}_{0}^{S}$ &  &  &  &  \\ 
$\Sigma _{1}={\cal A}_{1}^{F}$ &  &  &  &  \\ 
$\Sigma _{2}=-8~{\cal J}_{2}^{V}-{\cal J}_{2}^{F}+{\cal J}_{2}^{S}$ &  & $%
\Xi _{2}=\frac{4}{3}~{\cal J}_{2}^{V}+\frac{4}{3}~{\cal J}_{2}^{F}+{\cal J}%
_{2}^{S}$ &  &  \\ 
$\Sigma _{3}=-8~{\cal A}_{3}^{V}+{\cal A}_{3}^{F}$ &  & $\Xi _{3}=\frac{16}{5%
}~{\cal A}_{3}^{V}+{\cal A}_{3}^{F}$ &  &  \\ 
$\Sigma _{4}=8~{\cal J}_{4}^{V}-2~{\cal J}_{4}^{F}+{\cal J}_{4}^{S}$ &  & $%
\Xi _{4}=-\frac{72}{5}~{\cal J}_{4}^{V}-\frac{3}{5}~{\cal J}_{4}^{F}+{\cal J}%
_{4}^{S}~$ &  & $\Upsilon _{4}=\frac{16}{5}~{\cal J}_{4}^{V}+\frac{8}{5}~%
{\cal J}_{4}^{F}+{\cal J}_{4}^{S} $ \\ 
&  & $\Xi _{5}=-8~{\cal A}_{5}^{V}+{\cal A}_{5}^{F}$ &  & $\Upsilon _{5}=%
\frac{32}{7}~{\cal A}_{5}^{V}+{\cal A}_{5}^{F}$ \\ 
&  & $\Xi _{6}=8~{\cal J}_{6}^{V}-2~{\cal J}_{6}^{F}+{\cal J}_{6}^{S}$ &  & $%
\Upsilon _{6}=-\frac{120}{7}~{\cal J}_{6}^{V}-\frac{3}{7}~{\cal J}_{6}^{F}+%
{\cal J}_{6}^{S}$ \\ 
&  &  &  & $\Upsilon _{7}=-8~{\cal A}_{7}^{V}+{\cal A}_{7}^{F}$ \\ 
&  &  &  & $\Upsilon _{8}=8~{\cal J}_{8}^{V}-2~{\cal J}_{8}^{F}+{\cal J}%
_{8}^{S}$%
\end{tabular}
\]

This table (that can continue for arbitrarily high spin) is constructed by
combining N=4 supersymmetry, in particular the $\delta_\zeta^2$ operation,
and orthogonality. The former rule dictates the vertical movement in the
table (OPE-singularity direction), the latter prescribes the horizontal one
(spin of the lowest component). For example, at each odd-spin $2s-1>1$ there
are two orthogonal currents. Supersymmetry determines their two spin-$2s$
partners. Then orthogonality determines the third spin-$2s$ current.
Afterwards supersymmetry generates two spin-$(2s+1)$ currents (one spin-$2s$
current being annihilated) and so on.

As a first check of the above table, one can prove that $<\Sigma
_{4}~\Xi _{4}>=0$, using the results of \cite{high} for the two-point
functions. Indeed, factorizing out the common factor $N_{V}/16\pi ^{4}~{%
\prod }^{(4)}\left( 1/|x|^{4}\right) $, the tables of ref.\ \cite{high}
give: $<{\cal J}_{4}^{S}~{\cal J}_{4}^{S}>=32/105,$ $<{\cal J}_{4}^{F}~{\cal %
J}_{4}^{F}>=8/63$ and $<{\cal J}_{4}^{V}~{\cal J}_{4}^{V}>=1/252.$ $\Upsilon
_{4}$ is then determined by imposing the orthogonality conditions $<\Sigma
_{4}~\Upsilon _{4}>=<\Xi _{4}~\Upsilon _{4}>=0$.

A second check is $<\Sigma _{5}~\Xi _{5}>=0$. Indeed according to \cite
{high} $<{\cal A}_{5}^{F}~{\cal A}_{5}^{F}>=64/1155$ and $<{\cal A}_{5}^{V}~%
{\cal A}_{5}^{V}>=1/660$, factorizing out the common factor $N_{V}/16\pi
^{4}~{\prod }^{(5)}\left( 1/|x|^{4}\right) .$

It is important to observe that this table can be worked out at the
free-field level. To continue it, one would need to generalise the formulas
of \cite{high} to arbitrary spin (and we hope that our results stimulate the
study of this algebraic problem). There is a third direction
that can be added to the table, namely
turning the interactions on (anomalous dimension), described in Fig. 1. 
It would be interesting to work out the complete spectrum.
The problem is still algebraic to first order and 
the general formula will 
be presumably rather simple, due 
to the ``prime-factor-rule''. It should be similar to those
that one finds in the context of deep inelastic scattering.

We now observe that $\Xi _{5,6}$ have the same form as $\Sigma _{3,4}$ and $%
\Upsilon _{7,8}$, but different spin. The special combination $8~{\cal J}%
^{V}-2~{\cal J}^{F}+{\cal J}^{S}$ is the {\it only one} which is annihilated
by the $\delta _{\zeta }^{2}$ operation and therefore it has to be the
lowest term of any finite current multiplet. It can come only from the
combination $-8~{\cal A}^{V}+{\cal A}^{F}$, which, in turn, can come from
many. Precisely, the square of the $\delta_\zeta^2$ operation acts on $\{ 
{\cal A}^V,{\cal A}^F\}$ via the matrix $N=\left(\matrix{1 & 8\cr
-1/8 & -1}\right)$ and therefore produces $-8~{\cal A}^{V}+{\cal A}^{F}$ on 
any linear combination of ${\cal A}^{V}$ and ${\cal A}^{F}$. We
conclude that the structure of the current multiplets is universal, with a
rational two-parameter freedom. 
Any current multiplet is 4-spin long, apart from the
stress-tensor multiplet, and has the form 
\begin{eqnarray}
\Lambda _{2s}&=&a_s~{\cal J}_{2s}^{V}+b_s~{\cal J}_{2s}^{F}+c_s~{\cal J}%
_{2s}^{S} \qquad \rightarrow \qquad  \nonumber \\
\Lambda_{2s+1}&=&-16(a_s+4b_s)~{\cal A}_{2s+1}^{V}+ (a_s-8b_s-24c_s)~{\cal A}%
_{2s+1}^{F} \qquad \rightarrow  \nonumber \\
\Lambda_{2s+2}&=&24(a_s+8b_s+8c_s)~{\cal J}_{2s+2}^{V}- 3(a_s-8c_s)~{\cal J}%
_{2s+2}^{F}+(a_s-8b_s-24c_s)~{\cal J}_{2s+2}^{S} \qquad \rightarrow 
\nonumber \\
\Lambda_{2s+3}&=&-8~{\cal A}_{2s+3}^{V}+{\cal A}_{2s+3}^{F} \qquad
\rightarrow \qquad  \nonumber \\
\Lambda _{2s+4}&=&8~{\cal J}_{2s+4}^{V}-2~{\cal J}_{2s+4}^{F}+{\cal J}%
_{2s+4}^{S}~~.
\end{eqnarray}
The two-parameter rational set of currents can
be represented as a discrete set of points on a
plane: if $\Lambda_s$ is 
the highest component of a current
multiplet, a generic point on the plane $(%
{\cal J}^V,{\cal J}^F)$, then the transformation $\Lambda_{2s}\rightarrow
\Lambda_{2s+2}$ is the projection onto the line $y=-x/16-3/2$, while $%
\Lambda_{2s+2}\rightarrow \Lambda_{2s+4}$ is the projection of this line
onto the ``fixed point'' $(8,-2)$ (see Fig. 2).

\let\picnaturalsize=N

\ifx\nopictures Y\else{\ifx\epsfloaded Y\else\fi
\global\let\epsfloaded=Y \centerline{\ifx\picnaturalsize N\epsfxsize
3.0in\fi \epsfbox{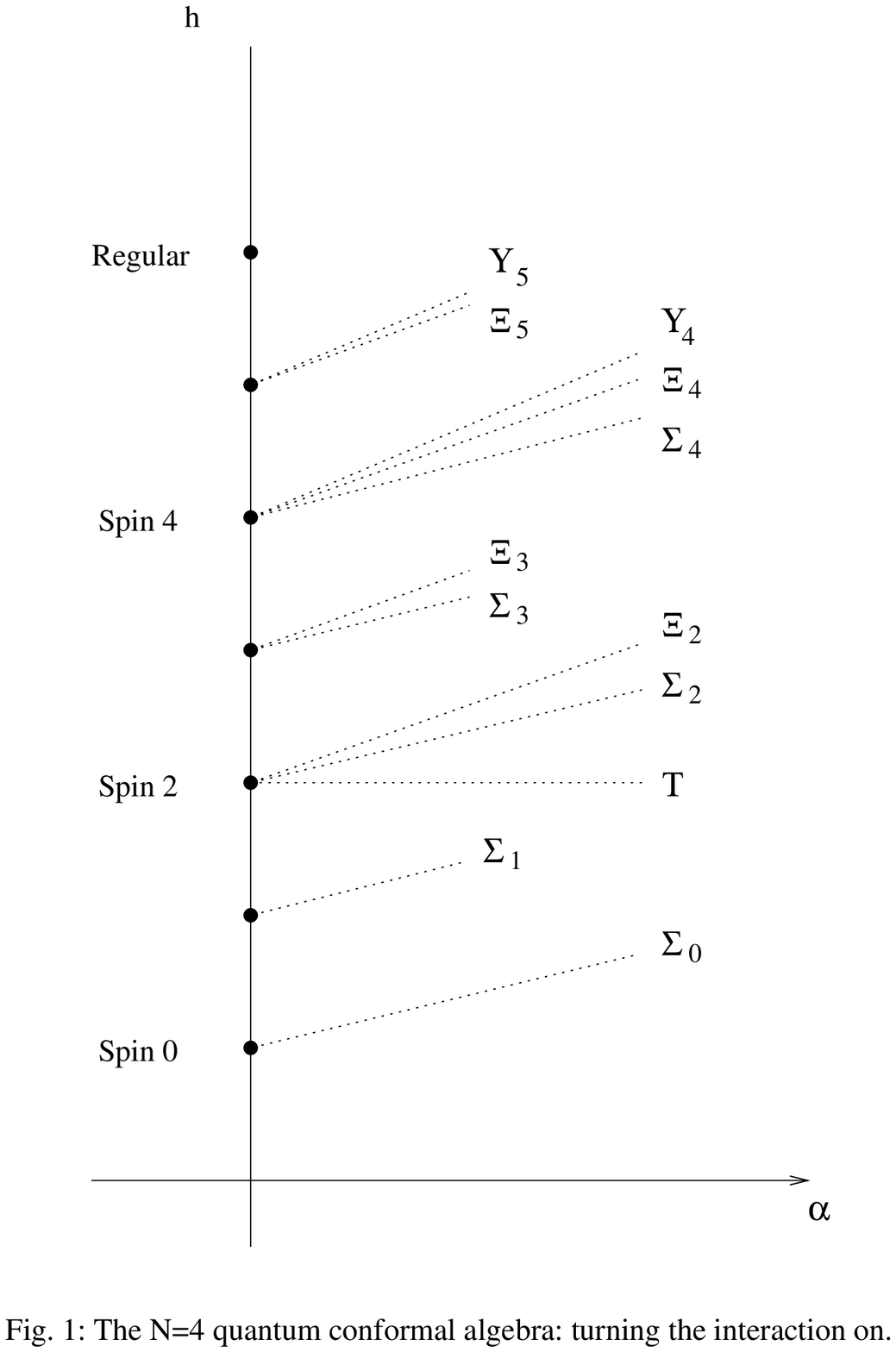}}}\fi
 
\let\picnaturalsize=N

\ifx\nopictures Y\else{\ifx\epsfloaded Y\else\fi
\global\let\epsfloaded=Y \centerline{\ifx\picnaturalsize N\epsfxsize
6.0in\fi \epsfbox{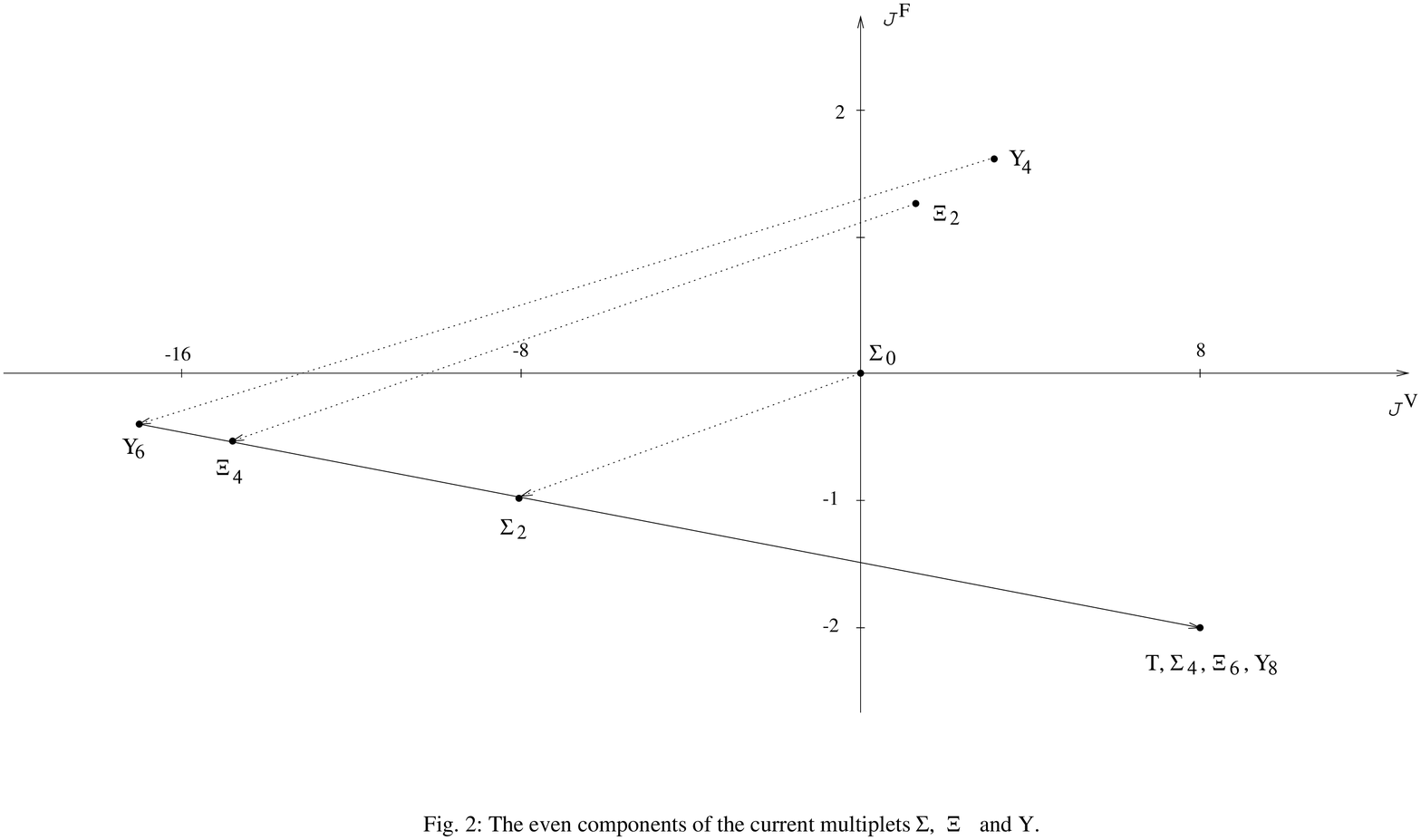}}}\fi

Via the procedure that we have outlined, the $TT$ OPE
determines algorithmically the set 
$(T,\Lambda_{2s})$, $s=0,1,\ldots$,
formed by the stress-tensor and one 4-spin-long 
current multiplet for each even spin.
$a_{s}$, $b_{s}$ and $c_{s}$
appear to involve prime factors 
not larger than $2s+1$ (see \cite{high} for this
number theoretical relationship). Moreover Fig. 2 suggests that,
fixing the coefficient of ${\cal J}^S$ to one,
all the points $(x_s,y_s)$ on the $({\cal J}^V,{\cal J}^F)$ plane
are located to the above-left of the stress-tensor, i.e.
$x_s\leq 8$ and $y_s\geq -2$ $\forall s$.

The currents of the set $(T,\Lambda_{2s})$ 
are a basis. All the $SU(4)$-invariant
currents with a nonvanishing free-field limit,
quadratic in the fields, can be expressed in terms of $(T,\Lambda_{2s})$.
The singular terms of the
OPE's of $\Sigma $, $\Xi $, $\Upsilon $ and all 
$\Lambda_{2s}$'s themselves can
be expressed via this basis.
Our construction determines
also the set $(c,c_{2s})$ of higher-spin central charges
defined in \cite{high}. 
The basis $(T,\Lambda_{2s})$ depends on the theory
and actually identifies its quantum conformal algebra. 

Summarizing, we have defined an algorithm to find
the currents of the quantum conformal algebra, starting from the
stress-tensor $T$ and the spin-0 component $\Sigma_0$ of the Konishi
multiplet and proceeding
via a combination of two operations:

i) supersymmetry, that moves ``vertically'' in the algebra; 

ii) orthogonalization of two-point functions, that moves ``horizontally''. 

\noindent In principle, 
all $\Lambda_{2s}$ can be determined with this procedure.
From the
practical point of view, however, the task might
be rather difficult: in order to work out, say, 
the currents of the multipelt $\Lambda_{100}$, 
one has to know all the currents with 
spin smaller than 100.
Some sort of ``hierarchy'' governs 
the mathematics of our construction. The problem is
to solve this hierarchy explicitly,
constructing a generating functional for the multiplets $\Lambda_{2s}$.
This goal might be achieved by combining
our procedure with techniques used
in the domain of higher-spin fields.
In particular, some aspects investigated in \cite{sezgin} have points
in common with our construction. In \cite{sezgin}
an infinite tower of higher-spin currents is generated by a 
product of two ``supersingletons''. This approach
might have applications to the question we are adressing
and reveal the 
intrinsic relationship between OPE's of ordinary
quantum field theory and higher-spin fields. Solving 
the quantum conformal hierarchies
is a challenging and intriguing problem.

We now perform the computation of the 
anomalous dimension of the $\Upsilon $ multiplet. We study the spin-4 vector
current 
\[
{\cal J}_{\mu \nu \rho \sigma }^{V}=\sum_{{\rm symm}}-4~D_{\mu }F_{\rho
\alpha }^{+}~D_{\nu }F_{\alpha \sigma }^{-}+\frac{6}{7}\partial _{\mu
}\partial _{\nu } \left( F_{\rho \alpha }^{+}F_{\alpha \sigma }^{-}\right) -%
{\rm traces.} 
\]
As before, we have written the terms in a form that does not contain double
covariant derivatives. The
expressions of $\Sigma_4$, $\Xi_4$ and $\Upsilon_4$ give 
\begin{equation}
{\cal J}^V_4={\frac{5}{1848}} (11 \,\Sigma_4 - 18 \,\Xi_4 + 7 \,\Upsilon_4).
\end{equation}
The form of the correlator of two operators ${\cal J}^{(i)}_4=\{
\Sigma_4,\Xi_4,\Upsilon_4\}$ is \cite{high} 
\[
<{\cal J}_4^{(i)}(x)\,{\cal J}_4^{(j)}(0)>=\left( \frac{1}{4\pi ^{2}}\right)
^{2}{\frac{c_{i}^{(4)}\delta _{ij}}{(|x|\mu )^{2h_{i}}}}{\prod }^{(4)}\left( 
{\frac{1}{|x|^{4}}}\right) =c_{i}^{(4)}\delta _{ij}\left( \frac{1}{4\pi ^{2}}%
\right) ^{2}{\frac{{\cal I}^{(4)}(x)}{(|x|\mu )^{2h_{i}}|x|^{12}}}, 
\]
where $\{c_i^{(4)}\}=\{16N_V/15,88N_V/75,352N_V/525\}$. The correlator of
the two divergences is 
\begin{equation}
<\partial {\cal J}_4^{(i)}(x)\,\partial {\cal J}_4^{(j)}(0)> =-{\frac{1800}{7%
}}h_{i}\left( \frac{1}{4\pi ^{2}}\right) ^{2}{\frac{c_{i}^{(4)}\delta _{ij}}{%
|x|^{4}}}{\prod }^{(3)}\left( {\frac{1}{|x|^{4}}}\right) ,  \label{div4}
\end{equation}
where-from we get 
\begin{equation}
<\partial {\cal J}_4^{V}(x)\,\partial {\cal J}_4^{V}(0)>=-{\frac{625}{332024}%
}\left( {\frac{1}{4\pi ^{2}}}\right) ^{2}{\frac{(121\,c^{(4)}_{\Sigma }h_{\Sigma
}+324\,c^{(4)}_{\Xi }h_{\Xi }+ 49\,c^{(4)}_{\Upsilon }h_{\Upsilon })}{|x|^{4}}}{\prod }%
^{(3)}\left( {\frac{1}{|x|^{4}}}\right) , 
\label{t1}
\end{equation}
The explicit computation gives
\begin{equation}
<\partial {\cal J}_4^{V}(x)\,\partial {\cal J}_4^{V}(0)>=-{\frac{590}{147}%
}\left( {\frac{1}{4\pi ^{2}}}\right) ^{3}{\frac{g^2N_cN_V}{|x|^{4}}}{\prod }%
^{(3)}\left( {\frac{1}{|x|^{4}}}\right).
\label{t2}
\end{equation}
The match of the full tensorial structure
appearing on the right hand side
is a severe check of the calculation.
In turn, it is also a confirmation that the higher-spin techniques
developed so far for conformal field theories work correctly.
The tensorial structure is uniquely fixed by conformality and
a finer check is obtained by observing that
the divergence  $\partial {\cal J}_4^{V}$ is the sum of two separately conformal terms
(a similar remark was made for spin 3), one linear in the field strength and $E^a_\mu$
and the other cubic in the field strength.

By comparison of (\ref{t1}) and (\ref{t2}) one obtains the final result
\[
h_\Upsilon={49\over 10}N_c{\alpha\over \pi}.
\]
A further check is provided by the ``prime-factor-rule'', according to which 
the basic algebraic quantities
should contain no prime factor higher than $2s+1$. At most, higher prime
factors can appear in the intermediate steps of the calculation. 
In our case, for example, the prime number 59 appearing 
in (\ref{t2}) disappears in the result. 

Other violations, relatively under control, come by taking
sums of algebraic quantities that obey the rule.
An example is provided by the higher-spin central 
charges $c^{(s)}_i$, where the prime factor $2s+3$ appears also
(check this for $s=2$ and $s=4$), despite each contribution
(scalar, spinor, vector) to the same quantities
obeys the $(2s+1)$-rule.

\vskip 1truecm

{\bf General properties at arbitrary coupling}.

\vskip .3truecm

In this section we describe the phenomena 
that happen when the valued of the coupling constant is not small.

Our results are consistent with certain
theorems that were worked out in the context
of the theory of deep inelastic scattering and the light-cone operator 
product expansions of electromagnetic and weak currents.
The Euclidean OPE differs considerably from the light-cone one,
in the sense that the terms are differently organized
(for example, infinitely many operators, including
their descendants, have the same light-cone singularity).
Nevertheless, 
the theorems we are talking about are completely general and follow
just from unitarity 
and dispersion relations.

Our anomalous dimensions are always positive.
This is in agreement with the theorem 
of Ferrara, Gatto and Grillo \cite{grillo}, 
which states that they cannot be negative,
as a consequence of unitarity.
Moreover, our values increase with the spin and the difference between
two consecutive values decreases with the spin.
According to a theorem by Nachtmann \cite{nach},
the lowest (most singular) value $h_{2s}$ of the 
anomalous dimensions of the operators of even spin $2s$, $s\geq 2$,
considered as a function of $s$, is increasing and convex:
\[
h_{2s}\leq h_{2(s+1)},~~~~~~~~~~~~~~~
h_{2(s+1)}-h_{2s}\leq h_{2s}-h_{2(s-1)}.
\]
The four values $\{h_T,h_\Sigma,h_\Xi,h_\Upsilon\}=\{
h_2,h_4,h_6,h_8\}
=\left\{0,3,{25\over 6},{49\over 10} 
\right\}\, N_c \, {\alpha\over \pi}$, do satisfy these properties,
illustrated in Fig. 1.

An important consequence is
that no conserved current
other than the stress-tensor appears 
in the N=4 quantum conformal algebra, at least generically,
i.e. apart from special points $g_*$ such that $h(g_*^2)=0$ $\forall s$.
Moreover, when the value of the
coupling constant increases
the multiplets move altogether. They never cross one another and
the relevant structure of the algebra
is the same at an arbitrary magnitude of the interaction:
no dramatic phenomenon can occur in the N=4 conformal family.

A richer set of phenomena does
occur in general (see \cite{high}, sect. 4.5, for an example
in the SQCD conformal window), but
for this to happen, one needs
more current multiplets with the same highest and lowest spins
(the Nachtmann theorem puts restrictions only to the 
minimal anomalous dimension of each even-spin level of the OPE).
The N=1 and N=2 quantum conformal algebras exhibit such
a richer set of phenomena, as we will show in 
\cite{n=2}.

A consequence of the Nachtmann theorem
is an alternative derivation the statement,
derived in the first section of this paper
using OPE considerations,
that the existence of a conserved current
with spin greater than 2 implies the existence of
infinitely many conserved currents (at least one
for each even spin greater than 2). 
Suppose $h_{2s}(g_*^2)=0$ for
some $s>1$ and some $g_*$. Now, $h_2$ is always zero for any $g$,
because of the presence of the stress tensor.
The only non-negative increasing convex function passing twice 
by zero is the null function $h_{2s}(g_*^2)\equiv 0$. 
In the N=4 case, all the currents
are conserved in these points $g_*$.

In summary, only two remarkable events can take place
in the N=4 
quantum conformal algebra, when varying the parametes $g$ and $N_c$: 
i) $h_{2s}=0$
$\forall s>0$
and ii) $h_{2s}=\infty$ $\forall s>0$. In the former case the theory is free, 
in the latter case the $TT$ OPE closes.

At $N_c$ fixed, $g\rightarrow {1\over g}$ duality arguments
suggest that $h_s(g^2)=h_s(1/g^2)$. $h_s$ should
have a maximum at $g\sim 1$ and all $h_s$'s should go back
to zero in the limit of infinitely strong coupling.
Instead, in the limit of large $N_c$
and $g^2N_c$,
the functions are expected to tend to infinity,
which means that the OPE should close solely
with the stress-tensor. This fact was conjectured in 
\cite{high}, sect. 4.5. 

The Nachtmann theorem does not apply before spin 2.
Therefore, there {\it can} be conserved 
spin-1 currents or even finite scalar operators without
having an infinite symmetry.
This does not happen in the case 
that we have considered in this paper, but it is
a property, for example, of the N=2 quantum conformal algebra
\cite{n=2}.

With these remarks, we think that we have determined
all the salient features of the N=4 quantum conformal algebra.

\vskip .5truecm

{\bf Conclusion.}

\vskip .3truecm

In this paper we have worked out the quantum conformal algebra of N=4 
super-symmetric Yang-Mills theory and used it as a technique
to study the N=4 conformal family at a generic magnitude of the interaction. 
Here we make some simple final comments.

As long as our knowledge deepens,
conformal field theories in four dimensions continue to exhibit,
at the same time,
similarities and differences with respect to conformal field
theories in two dimensions. The similarity discussed in \cite{high}
and applied here to a concrete model is the importance of higher spin currents.
The differences that emerge are the specific role 
played by higher spin currents
and their properties. 

In two dimensions, suitable perturbations
of conformal field theories create a mass gap
and, at the same time, preserve an infinite symmetry, precisely
an infinite set of
conserved higher spin currents. These properties 
make the model integrable and the $S$-matrix
exactly computable.
In four dimensions, instead, each (eventually
non-conserved) higher spin current
is associated with a perturbation of the conformal theory
and the anomalous dimension is the slope of the beta function
generated by this perturbation. For example, 
the multiplet of the stress tensor 
is associated with the marginal deformation (the N=4 lagrangian,
zero anomalous dimension),
the Konishi current multiplet $\Sigma$ is associated with the
N=1 deformation of the superpotential
\cite{noialtri}. Similar considerations apply to 
higher spin currents. It would be interesting 
to better identify the deformations associated with $\Xi$ and $\Upsilon$ and 
check the values $h_\Xi$ and $h_\Upsilon$ by computing
the slopes of the corresponding beta-functions.

We believe that these concepts are
interesting not only in an abstract sense, enriching our knowledge
of new algebraic structures and many related intriguing problems.
We expect that these ideas will help us getting a better description of 
critical points, which, in a long-range
perspective, can be useful both to quantum field theory,
hopefully QCD, and 
condensed matter physics, in particular (high temperature) super-conductivity.
Our plan is to use these techniques in order
to solve the conformal window of QCD.

\vskip .5truecm

{\bf Acknowledgements.}

\vskip .1truecm

I am grateful to S. Ferrara for discussions,
G. Mussardo for correspondence on two-dimensional
integrable models
and CERN for hospitality during the final stage of this work. 
This work is partially supported by EEC grants CHRX-CT93-0340 and TMR-516055.

\end{document}